\documentclass[aps,prl,preprint,groupedaddress]{revtex4-1}

\usepackage{graphicx}

\begin{document}


\title{On the Search for Time Reversal Invariance Violation in Neutron Transmission}


\author{J. David Bowman}
\email[]{bowmanjd@ornl.gov}
\affiliation{Oak Ridge National Laboratory, Oak Ridge, TN 37831}

\author{Vladimir Gudkov}
\email[]{gudkov@sc.edu}
\affiliation{Department of Physics and Astronomy, University of South Carolina, Columbia, SC 29208}


\date{\today}

\begin{abstract}
Time Reversal Invariant Violating (TRIV) effects in neutron transmission through a nuclei target are discussed. We demonstrate the existence of a class of experiments that are free from falls asymmetries.  We discuss enhancement of TRIV effects. We analyze a model experiment and show that such tests have a discovery potential of  $10^2 - 10^4$ compare to current limits.   
\end{abstract}


\maketitle

\section{Introduction}

Time reversal invariance violation (TRIV)
in nuclear physics has been studied for several decades. There are a number of TRIV effects in nuclear reactions and nuclear decays, which are sensitive to either
CP-odd and P-odd (or T - and P-violating) interactions or
T -violating P-conserving (C-odd and P-even) interactions. Here we consider   TRIV effects in nuclear reactions which can be measured in a transmission of polarized neutrons through a polarized target\cite{Kabir:1982tp,Stodolsky:1982tp}.
Such reactions can be described within the framework of neutron optics (for a discussion of neutron optics and see for example \cite{Gurevich,Squires}.)  The transmitted neutron wave propagates through a medium according to a spin-dependent index of refraction. The index of refraction depends on any applied magnetic field and the polarization of the medium. Because the state of the medium does not change, the polarization of the medium can be treated as a classical field. Because the initial and final propagation vectors of the neutron are the same, the initial and final states of the neutron can be time reversed by rotation the apparatus.
The neutron and nuclei are composite systems and may have accidetial cancelation of TRIV effects.
The important advantage in searching for TRIV in this process is the variety of  nuclear systems to measure T-violating parameters. This provides assurance  that  possible ``accidental'' cancelation of T-violating effects due to unknown structural factors related to the strong interactions in the particular system can be avoided.  Taking into account that different models of the CP-violation may contribute differently to a particular T/CP-observable, which  may have  unknown theoretical uncertainties, TRIV nuclear effects could be considered   complementary   to  electric dipole moment (EDM) measurements.
 Moreover,there is the possibility of an enhancement of T-violating observables  by many orders of  magnitude  due to the complex nuclear structure  (see, i.e. paper \cite{Gudkov:1991qg} and references therein).

For the observation of TRIV and parity violating (PV) effects,  one can consider  effects related to the T-odd correlation, $\vec{\sigma}_n\cdot({ \vec{k}}\times{\vec{I}})$, where  $\vec{\sigma}_n$ is the neutron spin, ${\vec{I}}$ is the target spin,
and $\vec{k}$ is the neutron momentum, which can be observed in the transmission of polarized neutrons through a target with a polarized nuclei.
 This correlation leads to a
difference  between the total neutron cross sections \cite{Stodolsky:1982tp} $\Delta\sigma_{\not{T}\not{P}}$  for $\vec{\sigma}_n$
parallel and anti-parallel to ${\vec{p}}\times{\vec{I}}$
and to neutron spin rotation angle \cite{Kabir:1982tp}  $\phi_{\not{T}\not{P}}$  around the axis
$\vec{p}\times{\vec{I}}$
\begin{equation}
\label{eq:dpt}
\Delta\sigma_{\not{T}\not{P}}=\frac{4\pi}{k}{\rm Im}(f_{\uparrow}-f_{\downarrow}),\qquad \frac{d\phi_{\not{T}\not{P}}}{dz}=-\frac{2\pi N}{k}{\rm Re}(f_{\uparrow}-f_{\downarrow}).
\end{equation}
Here, $f_{\uparrow,\downarrow}$ are the zero-angle scattering amplitudes for neutrons polarized
parallel and anti-parallel to the $\vec{k}\times{\vec{I}}$ axis, respectively;
 $z$ is the target length and $N$ is the number of target nuclei per
unit volume.
These TRIV effects  can be  enhanced \cite{Bunakov:1982is} by a factor of about $10^6$, and the similar  enhancement was observed for PV effects related to  $(\vec{\sigma}_n\cdot{ \vec{k}})$ correlation in neutron transmission through nuclear targets. For example, the PV asymmetry in the $.734 \ eV$ resonance in $^{139}La$ has been measured to be $(9.56 \pm 0.35)\cdot 10^{-2}$ (see, for example \cite{Mitchell:1999zz} and references therein).

  The PV and
TRI-conserving difference of total cross sections $\Delta
\sigma_{\not{P}}$ in the transmission of polarized neutrons through
unpolarized target which is proportional to the correlation
$(\vec{\sigma}\cdot \vec{k})$ can be written in terms of
differences of zero angle scattering elastic amplitudes with negative  and
positive  neutron helicities as:

\begin{equation}
 \Delta \sigma_P = {{4\pi}\over k}{\it Im}(f_- - f_+).
 \label{eq:dp}
\end{equation}

Then, one can calculate both TRIV and PV  amplitudes using distorted wave Born
approximation in the first power of parity and time reversal
violating interactions (see, for example ref.\cite{Bunakov:1982is}).  Thus,
the symmetry violating amplitudes can be written as
\begin{equation}
t^{fi}_{\not{P},\not{PT}} = <{\Psi^-_f}|V_{\not{P},\not{PT}}|{\Psi^+_i}>,
\end{equation}

 where $\Psi^{\pm}_{i,f}$ are the eigenfunctions of the nuclear T-invariant
 Hamiltonian with  the  appropriate boundary conditions \cite{Mahaux:233259}:
\begin{equation}
 \Psi^{\pm}_{i,f}=\sum_k a^\pm_{k(i,f)}(E)\; \phi_k + \sum_m\int
b^{\pm}_{m(i,f)}(E,E')\; \chi^{\pm}_m(E')\; dE'.
 \label{eq:wf}
\end{equation}
Here $\phi_k$ is the wave function of  the $k^{th}$
 compound-resonance and $\chi^{\pm}_m(E)$ is the potential
 scattering wave function in the channel $m$.
The coefficient
\begin{equation}
 a^\pm_{k(i,f)}(E)={\exp{(\pm i\delta_{i,f})}\over {(2\pi)^{1\over 2}}}{{(\Gamma^{i,f}_k)^{1\over 2}}\over {E-E_k\pm{i\over
   2}\Gamma_k}}
\end{equation}
describes compound nuclear resonances reactions and the
coefficient $b^{\pm}_{m(i,f)}(E,E')$ describes potential
scattering and interactions between the continuous spectrum and
compound resonances. (Here $E_k$, $\Gamma_k$, and $\Gamma^i_k$ are
the energy, the total width, and the partial width in the channel
$i$ of the $k$-th nuclear compound resonance, $E$ is the neutron
energy, and $\delta_i$ is the potential scattering phase in the
channel $i$; $(\Gamma^i_k)^{1\over 2} = (2\pi )^{1\over 2}
 <{\chi_i(E)}|V|{\phi_k}>$,
 where $V$ is a residual interaction operator.)

Since the dominant mechanism of symmetry violation
 in heavy nuclei is the mechanism of symmetry mixing on the compound
nuclear stage \cite{Bunakov:1982is}, only  first
term in Eq.~(\ref{eq:wf}) is important. For sake of simplicity we
consider the  case of a two resonance approximation, which is
 reasonably good for many heavy nuclei in the low
neutron energy  region $E \sim 1 eV - 100 eV$. Then, symmetry violating amplitudes due to mixing of the nearest $s$-wave and $p$-wave resonances could be written as:

\begin{equation}
<{p}|t|{s}> = - {1\over{2\pi}}
{{(v+iw)(\Gamma^n_s\Gamma^f_p)^{1\over
2}}\over{(E-E_s+i\Gamma_s/2)(E-E_p+i\Gamma_p/2)}}{\it
e}^{i(\delta^n_s + \delta^n_p)}, \label{eq:ps}
\end{equation}
and
\begin{equation}
<{s}|t|{p}> = - {1\over{2\pi}}
{{(v-iw)(\Gamma^n_p\Gamma^n_s)^{1\over
2}}\over{(E-E_s+i\Gamma_s/2)(E-E_p+i\Gamma_p/2)}}{\it
e}^{i(\delta^n_p + \delta^n_s)},
 \label{eq:sp}
\end{equation}

where $v$ and $w$ are real  matrix elements for PV and TRIV mixing between
$s$- and $p$-wave compound resonances
\begin{equation}
 v+iw = <{\phi_p}|V_{P}+V_{PT}|{\phi_s}>
  \label{eq:me}
\end{equation}
due to   $V_{\not{P}}$ (PV) and $V_{\not{P}\not{T}}$ (TRIV) interactions.
 One can see that PV and TRIV  matrix elements are  real and imaginary parts of  the same   matrix element calculated with
exactly the same wave functions. Also,
 the difference of amplitudes $(f_-
- f_+)$ for PV effect in Eq.~(\ref{eq:dp}) is
proportional to the sum of the symmetry violating amplitudes
(Eq.~(\ref{eq:ps}) and Eq.~(\ref{eq:sp})) but the difference of
amplitudes $(f_{\uparrow} - f_{\downarrow})$ for $PT$-violating
effect in Eq.~(\ref{eq:dpt}) is proportional to the difference of
the same amplitudes (Eq.~(\ref{eq:ps}) and Eq.~(\ref{eq:sp})).
This results in the same energy dependencies for both PV and
TRIV effects. Indeed, taking into account all numerical
factors one gets:
\begin{equation}
 \Delta\sigma_{\not{P}\not{T}} = - {{2\pi G_J}\over
k^2}{{w(\Gamma^n_s\Gamma^n_p(S))^{1\over 2}}\over{[s][p]}}[(E -
E_s)\Gamma_p + (E - E_p)\Gamma_s],
\end{equation}

and
\begin{equation}
 \Delta\sigma_{\not{P}} = {{2\pi G_0}\over
k^2}{{w(\Gamma^n_s\Gamma^n_p)^{1\over 2}}\over{[s][p]}}[(E -
E_s)\Gamma_p + (E - E_p)\Gamma_s],
\end{equation}
where  $[s,p]=(E-E_{s,p})^2+{{\Gamma^2_{s,p}}/4}$, $G_J$ and $G_0$
are spin factors; $J$ and $S$ are compound nuclei and channel
spins (see details in ref.\cite{Gudkov:1987re,Bunakov:1982is,Gudkov:1991qg}).
One can see that due to
similarity in these two equations,  the
TRIV effect has the same  enhancement as the
PV one.

 Now one can find the relation between the values of the PV and TRIV effects as
 \begin{equation}\label{svyaz}
 \Delta\sigma_{\not{T}\not{P}} = \kappa (J){w\over{v}}\Delta\sigma_{\not{P}},
 \end{equation}

 where
 \begin{eqnarray}
   \kappa (I+1/2) &=& -{3\over{2^{3/2}}}\left( {{2I+1}\over{2I+3}}\right) ^{3/2} \left( {3\over{\sqrt{2I+3}}}\gamma - \sqrt{I} \right) ^{-1}, \nonumber \\
  \kappa (I-1/2) &=& -{3\over{2^{3/2}}}\left( {{2I+1}\over{2I-1}}\right) \left( {I\over{I+1}}\right) ^{1/2} \left( -{{I-1}\over{\sqrt{2I-1}}}{1\over{\gamma}} + \sqrt{I+1} \right) ^{-1}.
 \end{eqnarray}

Here $\gamma = [\Gamma^n_p(I+1/2)/\Gamma^n_p(I-1/2)]^{1/2}$ is the ratio of  the neutron  width  amplitudes  for  the  different channel spins. In general, the parameter $\gamma$ may  be obtained from $\gamma$-ray  angular  correlation measurements in neutron  capture reactions \cite{Sushkov:1985ng,Bunakov:1982is}. Using standard unitary transformation one can rewrite the parameter $\gamma$ for neutron spin ($j=l\pm 1/2$) representation scheme as
\begin{equation}\label{gammasp}
\gamma = \frac{-\sqrt{2}\Gamma^n_p(1/2)^{1/2}+\Gamma^n_p(3/2)^{1/2}}{\Gamma^n_p(1/2)^{1/2}+\sqrt{2}\Gamma^n_p(3/2)^{1/2}}.
\end{equation}
 One can see from eq.(\ref{svyaz}), that the larger values of the parameter $\kappa (J)$ leads to the increasing of the sensitivity of TRIV difference of total  cross sections compare to PV one. This gives the opportunity to enhance the sensitivity of TRIV experiments by choosing a proper target.

\section{Enhancement factors}

Let us recall the main features of the enhancement factors for TRIV and PV effecs using as an example
P-odd   difference $\Delta\sigma_{P}$ of total cross sections.
 The quantity $\Delta\sigma_{P}$ displays resonance peaks  near both $s$- and $p$-wave resonances increasing its value by a factor of $(D/\Gamma )^2$ with respect to the point between the resonances $(D=\mid E_s-E_p\mid )$. These peaks are caused by the resonance enhancement of the wave function amplitude in the region of the interaction. The physical meaning of the resonance enhancement is quite obvious from the estimates of the compound-system life-time. This life-time $\tau$ can be expressed as the additional time, over the reaction time without resonance process, that the neutron spends in the range of the nuclear interaction. In terms of a neutron scattering phase shift $\delta (E)$, one can write

 \begin{equation}\label{lt}
\tau = 2 {d\delta (E)\over{dE}},
 \end{equation}
where the resonance part of the phase shift for the $i$-th resonance is $\delta (E) \simeq - \arctan{((\Gamma_i/2)/(E-E_i))}$.
 In the resonance state, the particle remains within the nucleus for a longer time of the order of the resonance life time $\sim (1/\Gamma )$. Therefore, it is natural to expect an enhancement of symmetry violation proportional to the ratio of the resonance lifetime $(1/\Gamma)$ to the lifetime of compound- nucleus away from the resonance $(\Gamma /D^2)$, that is to $(D/\Gamma )^2$. \par
  Let us consider the ratio $P = {\Delta \sigma_P}/(2\sigma_{tot})$, where $\sigma_{tot}$ is the total cross section and consists of the $s$-resonance , $p$-resonance
and the potential scattering contributions.                             The quantity $\sigma _{tot}$ also displays a marked resonance peak in the vicinity of  $s$ -wave resonance, which compensates completely for the corresponding peak of the numerator $P$. Therefore, the quantity $P$ is not enhanced in the vicinity of the $s$-wave resonance and remains approximately on the same level as the value between the resonances.
In general, $\sigma_{tot}$ is dominated by the smooth background of the $s$-wave resonance and potential scattering cross section in the vicinity of the $p$-wave resonance, since $(kR) \ll  1$ ($R$ is the nuclear radius). Therefore, the resonance peak of $\Delta \sigma_P$ near the $p$-resonance is retained in the quantity $P$, which is enhanced here by a factor of $(D/\Gamma )^2$
\begin{equation}\label{Pp}
P(E_p) \sim  8 {v\over D} \sqrt{{\Gamma^n_p}\over{\Gamma^n_s}} {{D^2}\over{\Gamma_s\Gamma_p}}  \left[ 1 + {{\sigma_p + \sigma_{pot}}\over{\sigma_s}}\right] ^{-1}.
\end{equation}
Here  $E_i$, $\Gamma_i$ and $\Gamma^n_i$ are the energy, total width and neutron width of the $i^{th}$ compound resonance, and $v$ is the nuclear weak matrix element.
 The presence of the penetration factor $\sqrt{{\Gamma^n_p}/{\Gamma^n_s}} \sim  (kR)$ in eq.(\ref{Pp}) is characteristic of all correlations observed in low energy nuclear reactions which arise due to initial state interference and, consequently, are proportional to the neutron momentum in the correlation $({\vec{ \sigma} \vec{ k}})$.
  It should be noted that $P$ might have the maximal magnitude
  \begin{equation}\label{Pmax}
   P_{max} \simeq  {{v\over D} {D\over {\Gamma }}} = {v\over{\Gamma}},
  \end{equation}

 when the total cross section contributions from the $s$- and $p$-resonances have similar magnitudes in the vicinity of the $p$-wave resonance.
 In addition to the resonance enhancement factor, there is the so called ^^ ^^ dynamic'' enhancement factor, which is connected with the ratio $v/D$. For a crude estimate of this ratio, one can expand the compound resonance wave function $\phi$ in terms of simple-configuration wave functions (e.g., one-particle wave functions)  $\psi_i$ which are admixed to compound resonances by strong interactions:
 \begin{equation}\label{eq}
\phi = \sum_{i=1}^N c_i\psi_i.
 \end{equation}

Using the normalization condition for the coefficients $c_i$ and the statistical  random-phase hypothesis for matrix elements $<{\psi_i}|W|{\psi_k}>$ we obtain
\begin{equation}\label{edq}
v = <{\phi_s}|W|{\phi_p}> = <{\psi_i}|W|{\psi_k}>_{RMS} N^{-1/2}.
\end{equation}
 Here $<{\psi_i}|W|{\psi_k}>_{RMS}$ is the root mean square  value of the matrix elements between simple configurations. In the black-nucleus statistical model, the number of components $N$ is estimated in terms of the average spacing $\overline{D}$ of compound resonances and the average spacing $\overline{D}_0$ of single-particle states:
 \begin{equation}\label{eq}
N \approx  \overline{D}_0/\overline{D}.
 \end{equation}

One can estimate $N$ from the experimental data on neutron strength functions since, in the statistical model of heavy nuclei, the neutron strength function is proportional to $N^{-1}$ (see, e.g., \cite{bohr1998nuclear}). The value of $N$ is about $10^6$. Hence
\begin{equation}\label{eq}
{v\over D} \simeq { {<{\psi_i}|W|{\psi_k}>_{RMS} }\over{\overline{D}_0}} \sqrt{N},
\end{equation}
where the ratio of the ^^ ^^ simple'' weak matrix element to the single particle level distance is about $10^{-7}$ (or the usual scale of the nucleon weak interaction). The enhancement factor $\sqrt{N}$ occurs as a result of the small level distance between nuclear compound resonances $(D^{-1} \sim  N)$ and the random-phase averaging procedure $( \sim  N^{-1/2})$.

Using the one particle formula (\ref{edq}) for the weak matrix element:
\begin{equation}\label{matv}
v \simeq  2\cdot 10^{-4} \sqrt{\overline{D}({\it eV})},
\end{equation}
 one can see that the maximal possible $P$ effect might be
 \begin{equation}\label{leq}
P_{max} \sim  10^{-4} \sqrt{\overline{D}({\it eV})}/\Gamma \leq  10\%
 \end{equation}
 for the case of medium and heavy nuclei, which  have typical values of the parameters $\overline{D} \in  (1 - 10^3){\it eV}$, $\Gamma  \in  (0.05 - 0.2){\it eV}$.

Using  one particle PV and
TRIV potentials
\begin{equation}
V_P = {G\over{8^{1/2}M}} \{ ({\vec{ \sigma}\cdot \vec{p}}),\rho
(\vec{r})\} _+,
\end{equation}

\begin{equation}
 V_{PT} = {{iG\lambda }\over{8^{1/2}M}} \{ ({\vec{\sigma}\cdot
\vec{p}}),\rho (\vec{r})\}_-,
\end{equation}
where $G$ is the weak
interaction Fermi constant, $M$ is the proton mass, $\rho (\vec{
r})$ is the nucleon density, $\vec{p}$ is the momentum of the
valence nucleon,
one can get a relation between   the ratio of
matrix elements $<\lambda
> = {w/v}$ and the ratio of nucleon coupling constants $\lambda = {g_{\not{P}\not{T}}/{g_{\not{P}}}}$ :
\begin{equation}
<\lambda > = {\lambda\over{1+2\xi}}.
\end{equation}
 Here $\xi \sim (1 - 7)$ (for detailed discussions see papers\cite{Gudkov:1990tb,Towner:1994qe,Khriplovich:1995zg,Desplanques:1995}).
with
\begin{equation}\label{ksi}
\xi = {{{<{\phi_p}|\rho ({\vec{\sigma}{\vec p}})|{\phi_s}>}\over{<{\phi_p}|{ ({\vec{\sigma}{\vec p}})}\rho |{\phi_s}>}}}.
\end{equation}
 $\phi_{s,p}$ are the $s,p-$resonance wave functions of the compound nucleus.
  The value of the matrix element in numerator can be estimated \cite{Gudkov:1990tb} using the operator identity $2{\bf p}=iM[H,{\bf r}]$ as
   \begin{equation}\label{num}
    {<{\phi_p}|\rho ({\vec{ \sigma} \vec{p}})|{\phi_s}>} \simeq  {{i\overline{\rho}M}\over 2}D_{sp}{<{\phi_p}|({\vec{ \sigma} \vec{p}})|{\phi_s}>}.
   \end{equation}
Here $H$ is the single particle nuclear Hamiltonian, $D_{sp}$ is the average single particle level spacing, and $\overline{\rho}$ is the average value of  the nuclear density. For denominator of eq.(\ref{ksi})   one can show
\begin{equation}\label{eq}
<{\phi_p}|{\bf (\sigma p)}\rho |{\phi_s}> =  - <{\phi_p}|{\bf (\sigma r)}{1\over{r}}{{\frac{\partial\rho}{\partial r}}}|{\phi_s}>
 =  {{2i\overline{\rho}}\over{R^2}}<{\phi_p}|{\bf (\sigma r)} |{\phi_s}>,
\end{equation}
where $R$ is the nuclear radius.
Then,  we obtain
  \begin{equation}\label{ksies}
  \xi  = {1\over{4}}MD_{sp}R^2 = {1\over{4}}\pi (KR),
  \end{equation}
where
\begin{equation}\label{eq}
D_{sp} = {1\over{MR^2}}\pi KR,
\end{equation}
for square-well potential model \cite{bohr1998nuclear}, with $K$ is the nucleon momentum in the nucleus, was used.
This leads to a value of $\xi \simeq  1$. Taking into account that  theoretical predictions for $\lambda$ are varying from $10^{-2}$ to $10^{-10}$ for different models of the CP violation (see, for example, \cite{Gudkov:1995tp} and references therein), one can estimate a range of  possible values of the TRIV observable and relate a particular mechanism of the CP-violation to their values.

\section{Final State Interactions}

The unique feature of the considered TRIV effects in neutron nuclei scattering (as well as the similar effects  related to TRIV and parity conserving correlation $\vec{\sigma}_n\cdot({ \vec{k}}\times{\vec{I}})\cdot({ \vec{k}}\cdot{\vec{I}})$) is  \emph{the absence of false TRIV effects due to the final state interactions (FSI)} (see, for example \cite{Gudkov:1991qg} and references therein), because these effect  are related to   elastic scattering at a zero angle. The general theorem about the absence of FSI for TRIV effects in elastic scattering has been proved first by R. M. Ryndin \cite{Ryndin:fsi} (see, also \cite{Ryndin:1965,Ryndin:1969,Gudkov:1991qg,Kabir:1988ma}). Since this theorem is very important, we give a brief sketch of the proof for the case of the zero angle elastic scattering following \cite{Ryndin:fsi,Gudkov:1991qg}. It is well known that the T-odd angular correlations in scattering and in a particle decay are not sufficient to establish TRIV, i.e. they have non-zero values in any process with strong, electromagnetic, and weak interactions.
For example, the analyzing power in the scattering of polarized particles $\vec{\sigma}\cdot({ \vec{k}_i}\times{\vec{k}_f})$  is odd under time reversal, and is known to be $O(1)$ for many systems.
This is because TRI, unlike parity conservation, does not provide a constrain on amplitudes of any process, but rather relates two different processes: for example,  direct and inverse channels of reactions.
However, for the case when the process can be described in the first Born approximation, we can relate T-odd correlations to TRIV interactions. Indeed,  the unitarity condition for the scattering matrix in terms of the reaction matrix $T$, which is proportional to the scattering amplitude, can be written as \cite{Landau:3}
\begin{equation}\label{unit}
T^{\dag}-T=iTT^{\dag}
\end{equation}
The first Born approximation can be used when  the right side of the unitarity equation is much smaller than the left side, and results in hermitian $T$-matrix
\begin{equation}\label{herm}
<i|T|f>=<i|T^*|f>,
\end{equation}
which with TRI condition
\begin{equation}\label{tri}
<f|T|i>=<-i|T|-f>^*
\end{equation}
leads to the constraint on the $T$-matrix as
\begin{equation}\label{todd}
<f|T|i>=<-f|T|-i>^*.
\end{equation}
This condition forbids T-odd angular correlations, as is the case with the P-odd correlations when  parity is conserved. (Here the minus signs in matrix elements mean the opposite signs for particle spins and momenta in the corresponding states.) For the case of the zero angle elastic scattering, the initial and final states  coincide ($i=f$), and when combined with TRI condition (\ref{tri}), result in  Eq.(\ref{todd}) without the violation of unitarity (\ref{herm}). Therefore, in this case, FSI cannot mimic  T-odd correlations, which  originated from TRIV interactions. Therefore, an observation of a non-zero value of  TRIV effects in neutron transmission directly indicates  TRIV, exactly like in the case of neutron EDM \cite{Landau:1957}.

However, to measure TRIV effects for neutron propagation with simple changing of neutron and/or nucleus polarizations is unpractical since it requires to high precision control for too many parameters which can contribute to systematic errors (see, for example, \cite{Lamoreaux:1994nd,Masuda:1992}. The approach to eliminate this difficulties was suggested in \cite{Kabir:1988ma} (see also \cite{Skoy:1996,Masuda:1998zb}), which will be implemented and discussed later in this paper.

\section{TRIV transmission theorem}

The   problem related to possible  false effects which can arise from one or more of the following sources: imperfect alignment of polarizer, target and analyzer, differences in the polarizer and analyzer, inhomogeneity of the target medium, rotations of the neutron spin due to the holding field of a polarized target, and interaction of the neutron spin with the target spin (pseudo-magnetic field) \cite{Baryshevsky:1964,abragam1982nuclear}.
Masuda \cite{Masuda96,Masuda:1998zb,Masuda:2000}, and Serebrov \cite{Serebrov96}  have proposed experiments that involve adding additional spin flips to the basic polarizer and polarized target apparatus. The difficulty in these approaches is that each added spin flip increases the number of parameters needed to characterize the apparatus by three: two alignment angles and an analyzing power.
 Lamoreaux and Golub \cite{Lamoreaux:1994nd} argue that, ``\ldots it will be necessary to develop new methods to make very precise absolute measurements of the neutron-spin direction. It seems hopeless to devise a experiment that would convincingly measure TRIV in the presence of such a wide variety of potential sources of false effects.''

\begin{figure}[h]
\includegraphics[width=100mm]{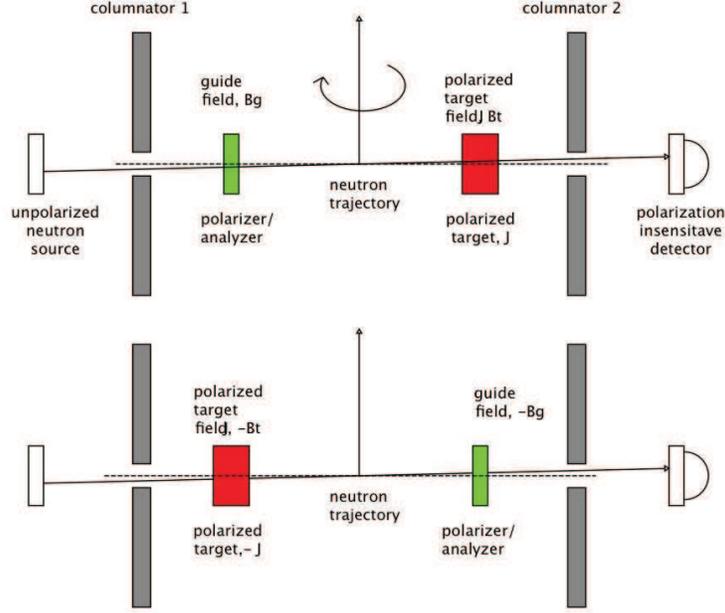}
\caption{ Schematic representation of an apparatus to measure TRIV in neutron optics.}
\label{apparatus}
\end{figure}

To resolve this problem we consider a configuration of the apparatus related to the approach originally proposed by Kabir \cite{Kabir:1988ma,Kabir:1989pc}, which is shown on Fig.(\ref{apparatus}),
where the polarizer and analyzer prepare and select spin perpendicular to neutron momentum $\vec{k}$. The target is polarized perpendicular to both  $\vec{k}$  and the polarizer direction.

To describe the transmission difference between these two shown configurations with both the polarizer and analyzer reversed, we can use the equation of motion for the neutron spin as the neutron propagates through a medium and an external magnetic field, $\vec{B}$,  given by Schrödinger's equation with the effective Hamiltonian (Fermi potential):
\begin{equation}\label{Ham}
H=\frac{2\pi \hbar ^2}{m_n}Nf-\frac{\mu}{2}(\vec{\sigma}\cdot \vec{B})
\end{equation}

where $m_n$  is the neutron mass, $n$  is the number of scattering centers per unit volume, $f$  is the forward elastic scattering amplitude, and $\vec{\sigma}$  are the Pauli spin matrices. (For discussion of the conditions under which equation (\ref{Ham}) applies, see \cite{Lamoreaux:1994nd} and references therein.) We can  write $f$  as the sum of four terms:
\begin{equation}\label{ampl}
f=a_0+b_0(\vec{\sigma}\cdot \vec{I})+c_0(\vec{\sigma}\cdot \vec{k})+d_0(\vec{\sigma}\cdot [\vec{k}\times \vec{I}]),
\end{equation}
where  $I$  is the polarization of a target medium, and quantities other than the neutron spin $\vec{\sigma}$ are treated as classical fields. Neutron spin-optics tests of TRIV have the goal of measuring $d$, which is the only term that originates from a TRIV interaction. Terms $a$  and $b$, give the strengths of the spin-independent, and strong spin-spin (pseudo magnetic) interactions, while terms $c$  and  $d$ give the degree of PV and TRIV arising from symmetry mixing in the neutron resonances in the target medium.

 We show that if all the classical fields that interact with the neutron spin are reversed, the forward and reversed transmissions for the apparatus configuration presented in Fig. \ref{apparatus} are equal if $d=0$. In the proposed approach, the magnetic field  $\vec{B}$ is reversed, and the orientations of   $\vec{I}$  with respect to $\vec{B}$  are maintained. Therefore, one can re-write Hamiltonian (\ref{Ham}) as
 \begin{equation}\label{HamEf}
H=a+b(\vec{\sigma}\cdot \vec{I})+c(\vec{\sigma}\cdot \vec{k})+d(\vec{\sigma}\cdot [\vec{k}\times \vec{I}]),
\end{equation}
 where $a=\frac{2\pi \hbar ^2}{m_n}N a_0$,  $b=\frac{2\pi \hbar ^2}{m_n}N b_0-(\mu B)/2$, $c=\frac{2\pi \hbar ^2}{m_n}N c_0$, and $d=\frac{2\pi \hbar ^2}{m_n}N d_0$.
 The acceptance of the apparatus is defined by a pair of collimators mounted on a rigid rotatable platform with the polarizers (analyzer) and target as it show in Fig.\ref{apparatus}. Rotating the apparatus  by an angle $\pi$  about an axis perpendicular to the symmetry axis of the collimators leaves fixed the family of accepted trajectories, but reverses the sign of $\vec{k}$  trajectory by trajectory. We assume that the product of the source strength and detector efficiency is symmetric with respect to the plane of the symmetry axis and the rotation axis. Then, the evolution operator for the forward neutron transmission, $U_F$, gives the relationship between the initial and final spin wave functions for a neutron trajectory that begins on the source, passes through the apparatus, and ends on the detector. Let us consider the case when we have only TRI interactions. Then we divide the apparatus into $m$ slabs and write the evolution operator $U_F$  as a time ordered product of the evolution operators for each of the slabs:
\begin{equation}\label{UF}
  U_F=\prod^{m}_{j=1}\exp{(-i\frac{\Delta t_j}{\hbar}H^F_j)}=\alpha +(\vec{\beta}\cdot \vec{\sigma}).
\end{equation}
Here $H^F_j$ is the Hamiltonian from equation (\ref{HamEf}) evaluated at slab $j$, and $\alpha$ and $\vec{\beta}$ contain only TRI terms, since TRIV parameter $d =0$.    In the expression for the reverse evolution operator, $U_R$, the time ordering of the product and the signs of the spin-dependent terms in $H^R_j$  are reversed from those in $H^F_j$. Then,  the reverse evolution operator is
\begin{equation}\label{UR}
  U_R=\prod^{1}_{j=m}\exp{(-i\frac{\Delta t_j}{\hbar}H^R_j)}=\alpha  -(\vec{\beta}\cdot \vec{\sigma}).
\end{equation}
The fact  that the signs of the spin-dependent terms in the evolution operator are reversed  leads to the possibility to eliminate systematic effects which may mimic TRIV effects in scattering experiments. This agrees with  Kabir's result about the possibility to unambiguously \cite{Kabir:1988ma} measure TRIV effects in neutron scattering.
Since the relation between forward and reverse evaluation operators is very important for further consideration and not obvious, we will prove it here.

First, let us consider two-slab medium. Then, the forward and reverse evaluation operators are
\begin{eqnarray}\label{U2}
 \nonumber
  U_F &=& U^{(1)}_F U^{(2)}_F=\exp{(-i\frac{\Delta t_1}{\hbar}H^F_1)}\exp{(-i\frac{\Delta t_2}{\hbar}H^F_2)}, \\
  U_R &=& U^{(2)}_R U^{(1)}_R=\exp{(-i\frac{\Delta t_2}{\hbar}H^R_2)}\exp{(-i\frac{\Delta t_1}{\hbar}H^R_1)}.
\end{eqnarray}

For the case of infinitesimally small widths of the slabs, each  exponential operator in the above equations can be written as
\begin{eqnarray}\label{Uslab}
 \nonumber
U^{(j)}_F & = & (1-i\frac{\Delta t_j}{\hbar}H^F_j)=F^{(j)} +  (\vec{A}^{(j)}\cdot \vec{\sigma}) , \\
 U^{(j)}_R &=& (1-i\frac{\Delta t_j}{\hbar}H^R_j) =  F^{(j)} -  (\vec{A}^{(j)}\cdot \vec{\sigma}) ,
\end{eqnarray}
correspondingly,
where
\begin{eqnarray}
\nonumber
  F^{(j)}&=& 1-i\frac{\Delta t_j}{\hbar}a^{(j)} ,\\
 \vec{ A}^{(j)} &=& \frac{-i\Delta t_j}{\hbar} (b^{(j)}  \vec{I} +c^{(j)}  \vec{k}) .
\end{eqnarray}
Therefore, these one slab evaluation operators have exactly the same structure as the operators in eqs. (\ref{UF}) and (\ref{UR}), provided $ F^{(j)} \rightarrow \alpha ^{(j)}$ and $\vec{ A}^{(j)} \rightarrow \vec{\beta ^{(j)}}$.
Then, the straight away calculations  for eq.(\ref{U2}) lead to exactly the same form as for eqs. (\ref{UF}) and (\ref{UR}), again, with
\begin{eqnarray}\label{U2r}
 \nonumber
 \alpha &=& {\alpha }^{(1)}{\alpha }^{(2)}+(\vec{\beta}^{(1)}\cdot \vec{\beta}^{(2)}),\\
 \vec{ \beta} &=& {\alpha }^{(1)}\vec{\beta}^{(2)}+{\alpha }^{(2)}\vec{\beta}^{(1)}- [\vec{\beta}^{(1)}\times \vec{\beta}^{(2)}].
\end{eqnarray}
Then, applying  mathematical induction, one can prove the proposition in general (multi-slab)  case as  is given in eqs. (\ref{UF}) and (\ref{UR}).
Now, applying this result for the calculations of the forward and reverse transmissions, $T_F$  and $T_R$ , for our experimental setup we obtain the relation
\begin{equation}\label{trans}
T_F=\frac{1}{2}Tr(U^{\dag}_F U_F)={\alpha}^*\alpha +(\vec{\beta}^*\vec{\beta})=\frac{1}{2}Tr(U^{\dag}_R U_R)=T_R,
\end{equation}
which we call TRIV transmission theorem.
  This theorem shows that if $d=0$ and whole apparatus is rotated   with the classical fields being reversed, then  the transmissions of (un-polarized) neutrons along the same trajectory in opposite directions are equal. Then, the sum of transmissions over the family of accepted trajectories is  equal, too. The proof of TRIV theorem makes no assumption concerning the geometrical symmetry of the classical fields and materials of the apparatus. Therefore, any deviation from the equality of the forward and reversed transmissions in eq.(\ref{trans}) is a clear manifestation of the existence of TRIV interactions (non-zero $d$ coefficient in eq.(\ref{HamEf})).
It should be noted that for non-zero $d$ coefficient the difference between $T_F$  and $T_R$ transmissions arises from both spin dependent and spin independent parts of the evolution operators, which is in agreement with Kabir's \cite{Kabir:1988ma,Kabir:1989pc} conclusion about the existence of a number of possible unambiguous tests.

\section{Evaluation of a model  experiment}

In spite of the discussed above advantages,  so far no TRIV experiment in neutron optics has been done: (1) until recently polarized targets of materials that have compound nuclear resonances that exhibit large PV asymmetries have not been available; and (2) it has proved difficult to devise an experiment that would eliminate false effects that arise from combinations of instrumental imperfections and TRI interactions of the neutron spin with materials and external fields.
In past years a considerable progress has been made on the first problem, when groups at the KEK national laboratory in Japan \cite{Masuda_91,Masuda_90} and Kyoto University \cite{Takahashi:1993np}, and at PSI in Switzerland \cite{Hautle:2000} have achieved substantial polarizations of $^{139}La$ nuclei in Lanthanum Aluminate crystals as large as $~10\ cc$. Thus $.734\ eV$ resonance in $^{139}La$, which has a longitudinal asymmetry of $9.5\%$ \cite{Yuan:1991zz}, is a good candidate for TRIV studies.

  As an example of practical implementation of the proposed experiment based on the TRIV transmission theorem, we can
	 use  cells of polarized $^3He$ as neutron polarizers and analyzers. The direction of the polarization of the $^3He$ polarization is always parallel to the magnetic field and reverses when the field direction is reversed adiabatically. It should be mentioned that ferromagnetic polarizers and analyzers can be difficult to use in this experiment because  hystersis effects prevent their precise reversal.  Also, since the earth's magnetic field can not be reversed,  it must be compensated  or shielded.  It is essential that the values of the classical fields be stable in time and magnetic field strengths and the polarizations of $^3He$ and the target medium can be accurately monitored using nuclear-magnetic-resonance techniques.

\begin{figure}[h]
\includegraphics[width=100mm]{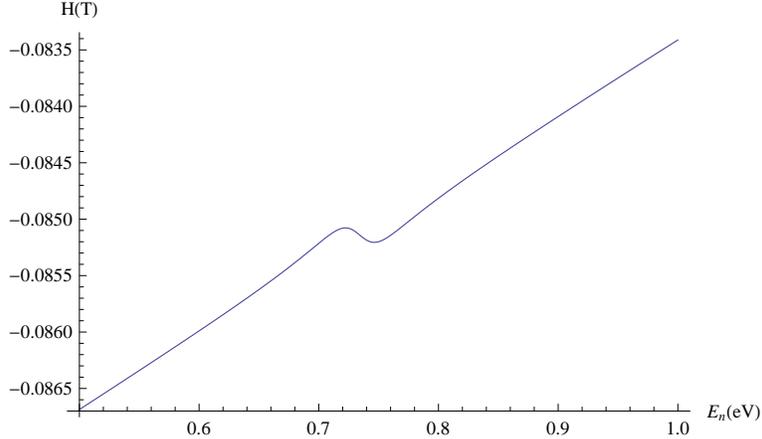}
\caption{ Pseudo-magnetic field in Lanthanum Aluminate crystals.}
\label{pseudoMag}
\end{figure}

For the target we can use $^{139}La$ nuclei in Lanthanum Aluminate crystals which has very large PV effect in the vicinity of $.734 \ eV$ resonance. Using experimentally archived value of  $^{139}La$ polarization of 47.5\%, we can estimate \cite{Baryshevsky:1964,Gudkov:1992pm} pseudo-magnetic field inside the crystal as a function of neutron energy (see Fig. \ref{pseudoMag}), which shows the pseudo-magnetic field is opposite the applied field. This gives an advantage for using Lanthanum Aluminate crystals, since values of TRIV effects in neutron optics, in general, are inverse proportional to the sum of magnetic and pseudo magnetic fields \cite{BG:FSI,Gudkov:1992pm}.

We make a rough estimate of the statistical uncertainty in the T-odd cross section that could be achieved in $10^7$ seconds of data collection on the water moderator of Flight Path 16A at the Spallation Neutron Source at Oak  Ridge National Laboratory. At the present time the Flight Path has not been instrumented. We assume a neutron production current of $1.4\ mA$ at $1\ GeV$ proton energy. We carry out the estimate for the $.734\ eV$ resonance in $^{139}La$. We assume that the target is one-interaction-length of dynamically polarized Lanthanum Aluminate and that the neutron beam is polarized by a one-interaction-length $70\%$ polarized $^3He$ spin filter.

We were unable to find a calculation or measurement of the neutron flux for FP16A. We estimated the neutron flux using the measurement of the flux from the water moderator of Flight Path 2 at the Los Alamos Neutron Scattering Center at the Los Alamos National Laboratory.  Roberson et al. \cite{Roberson:1993} found that the moderator brightness was well described by the expression
\begin{equation}\label{flux}
  \frac{ d^3N}{dA dt d \Omega}=k\frac{\Delta E}{E}\left( \frac{E}{1eV} \right)^{\gamma}\left( \frac{i}{e} \right)(neutrons\  cm^{-2} sec^{-1} sr^{-1}),
\end{equation}

with  $k=5.8 \cdot 10^{-3}$ and $ \gamma=0.1$. $E$  is the neutron energy, $i$   is the proton current, $e$  is the charge quantum, $A$  is the area of the moderator that is viewed, $\Delta E$   is the range of neutron energies accepted, and $\Omega$  is the solid angle acceptance of the apparatus. We assume that the neutron production rate is proportional to the proton energy and increase $k$  by 1000/800, the ratio of proton energies. We assume that SNS will operate at $1.4\ MW$ and  $i=1.4\ mW$.

We assume that  $A=100 cm^2$ and that the acceptance of the apparatus is defined by a $10\ cm$ diameter polarized target located 15 meters from the moderator: $\Omega =3.5\cdot 10^{-5}\ sr$. We set $ \Delta E= .045$ the total width of the resonance. The neutron flux is $dN/dt= 7.8\cdot 10^7\ neutrons/sec$.

In order to determine the uncertainty in the TRIV asymmetry we must make some assumptions concerning running time, source, polarizer, polarized target, detector, and cross sections. We assume a running time of $10^7\ sec$. We assume that the peak value of the resonance cross section is $2.9$ barns, the potential scattering cross section is $3.1$ barns, and the capture cross section at the resonance energy is $1.6$ barns. We assume the cross sections of aluminum and oxygen are  $3.8$ barns and $1.4$ barns \cite{Mughabghab}.  We calculate that the neutron polarization is $46\%$ and the transmission of the polarizer is $46\%$.  We assume a one-interaction-length $LaAlO_3$ target. We further reduce the transmission by a factor of 2 to account for various windows. The transmission of the apparatus is then estimated to be $11\%$. The transmitted beam intensity in $\Delta E$  is $Flux = .86\cdot 10^7 \ neutrons/sec$. The fractional uncertainty in TRIV cross section is given by
   \begin{equation}\label{frac}
   \frac{\delta\sigma}{\sigma}=\frac{1}{\sqrt{Flux \cdot Time}}\frac{\sum \sigma_k}{\sigma_p} .
 \end{equation}
(The sum runs over all the cross sections given above.) If we adopt the fractional parity-violating asymmetry for the resonance to be $9.5\%$ \cite{Yuan:1991zz}, we obtain an uncertainty in $\lambda$ , the ratio of the TRIV to PV asymmetries of $6.0 \cdot 10^{-6}$.

\section{Discovery Potential}

 Using the results of the recent calculations of PV and TRIV effects in neutron deuteron scattering \cite{Song:2010sz,Song:2011jh}, one can calculate the parameter $\lambda$ for this  reaction and  compare it to the case of the complex nuclei. Let us consider the ratio of the TRIV difference of total cross sections in  neutron deuteron scattering  given in \cite{Song:2011jh}
\begin{eqnarray}
\label{eq:PTP}
P_{\not{T}\not{P}}=\frac{\Delta\sigma_{\not{P}\not{T}}}{2\sigma_{tot}}&=&
\frac{(-0.185 \mbox{ b})}{2\sigma_{tot}}
[\bar{g}_\pi^{(0)}+0.26 \bar{g}_\pi^{(1)}
-0.0012 \bar{g}_\eta^{(0)}+0.0034 \bar{g}_\eta^{(1)} \\ \nonumber
&-&0.0071 \bar{g}_\rho^{(0)}+0.0035 \bar{g}_\rho^{(1)}
+0.0019 \bar{g}_\omega^{(0)}-0.00063 \bar{g}_\omega^{(1)}]
\end{eqnarray}
to the corresponding PV difference \cite{Song:2010sz}
\begin{eqnarray}
  \label{eq:PP}
P_{\not{P}}=\frac{\Delta\sigma_{\not{P}}}{2\sigma_{tot}}&=&
  \frac{(0.395 \mbox{ b})}{2\sigma_{tot}}[
  h_\pi^1  + h_\rho^0(0.021)+h_\rho^1(0.0027) \\ \nonumber
         &+&h_\omega^0(0.022)+h_\omega^1(-0.043)
         +h_\rho^{'1}(-0.012)
  ].
\end{eqnarray}
Here, we use one meson exchange model, known as the DDH model for PV nucleon interactions, to calculate both effects; in the above expressions, $\bar{g}$ and $h$ are meson- nucleon TRIV and PV coupling constants, correspondingly (see for details \cite{Song:2010sz,Song:2011jh}). From these expressions, one can see that  contributions from the pion exchange are dominant for both TRIV and PV parameters. Then, taking into account only the dominant pion meson contributions, one can estimate $\lambda$ as
\begin{equation}
\label{lambda}
\lambda=\frac{\Delta\sigma_{\not{T}\not{P}}}{\Delta\sigma_{\not{P}}} \simeq (-0.47)\left(\frac{\bar{g}^{(0)}_\pi}{h_\pi^1}
                   +(0.26)\frac{\bar{g}^{(1)}_\pi}{h_\pi^1}\right),
\end{equation}
which is in a good agreement with the estimate for the complex nuclei  \cite{Gudkov:1990tb}.

Also, we can relate the obtained parameter $\lambda$ to the existing experimental constrains obtained from EDM measurements, even though the relationships are highly model dependent.  For example, the CP-odd coupling constant $\bar{g}^{(0)}_\pi$ could be related to the value of the neutron EDM $d_n$ generated via a $\pi$-loop in the chiral limit \cite{Pospelov:2005pr}.  Then, using the experimental limit \cite{Baker:2006ts} on  $d_n$, one can estimate $\bar{g}^{(0)}_\pi $ as less than $2.5 \times 10^{-10}$. The constant $\bar{g}^{(1)}_\pi$ can be bounded using the constraint \cite{Romalis:2000mg} on the $^{199}Hg$ atomic EDM as $\bar{g}^{(1)}_\pi<0.5\times 10^{-11}$ \cite{Dmitriev:2003hs}.

The comparison of the $\lambda$ parameter  with the constrains on the coupling constants from the EDM experiments gives us the opportunity to estimate the possible sensitivity of TRIV effects to the value of TRIV nucleon coupling constant, which we call a ``\emph{discovery potential}'' for neutron scattering experiments \cite{Gudkov:2011tm,Gudkov:2013dp}, since it shows a possible factor for improving the current limits of the EDM experiments. Then, taking the DDH ``best value'' of $h_\pi^1\sim 4.6\cdot 10^{-7}$,  nuclear enhancement factors, and assuming that the parameter $\lambda$ could be measured with an accuracy of $10^{-5}$ on the complex nuclei, one can see from Eq.(\ref{lambda}) that  the existing limits on the TRIV coupling constants could be improved by  two orders of magnitude. It should be noted that to obtain Eq.(\ref{lambda}), the assumption was made that the $\pi$-meson exchange contribution is dominant for PV effects.  However, there is an indication \cite{Bowman} that the PV coupling constant  $h_\pi^1$ is much smaller than the ``best value'' of the DDH.  Should it be confirmed by the $\overrightarrow{n}+p \rightarrow d+\gamma$ experiment, the  estimate for the sensitivity of $\lambda$ to the TRIV coupling constant may be increased up to two orders of magnitude, as can be seen from Eqs.(\ref{eq:PTP}-\ref{lambda}). This  might increase the relative values of TRIV effects by two orders of magnitude, and as a consequence, the discovery potential of the TRIV experiments could be about $10^4$. Therefore, the TRIV effects in neutron transmission through a nuclei target are  unique TRIV observables being free from FSI, and are of the same quality as the EDM experiments. These TRIV effects are enhanced by about $10^6$ due to the nuclear enhancement factor. In addition to this enhancement, the sensitivity to TRIV interactions in these effects  might be  structurally enhanced by about $10^2$ if PV $\pi$-nucleon coupling constant is less than  the ``best value'' DDH estimate. Therefore, these types of experiments with high intensity neutron sources have a discovery potential of about $10^2 - 10^4$ for the improvement of the current limits on the TRIV interaction obtained from the EDM experiments.

\begin{figure}
\includegraphics[width=100mm]{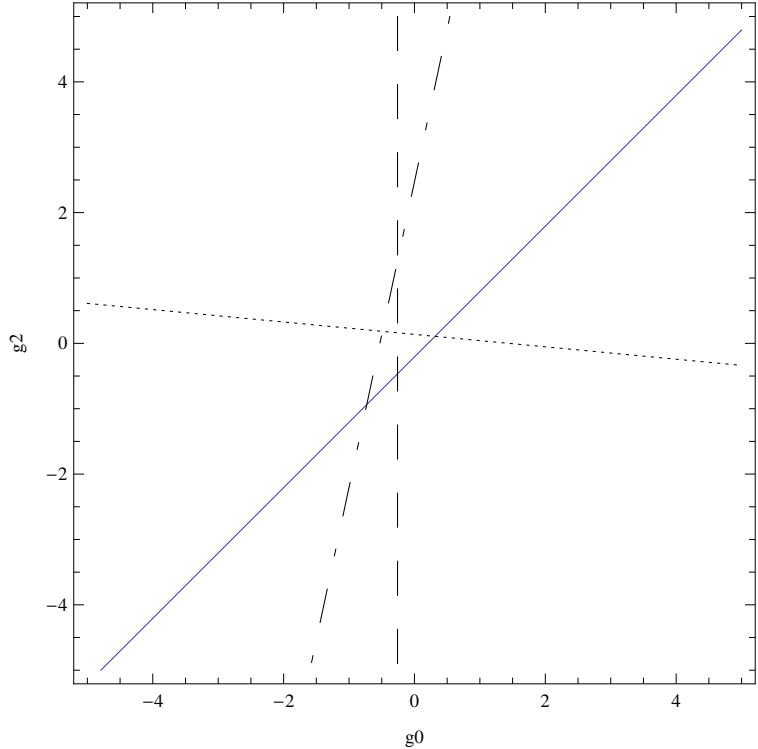}
\caption{The dependence of neutron EDM (solid), $^3He$ EDM (doted-dashed), $^3H$ EDM (doted) and parameter $\lambda$ on TRIV $\pi$-mesons iso-scalar and iso-tensor coupling constants.}
\label{iso02}
\end{figure}

\begin{figure}
\includegraphics[width=100mm]{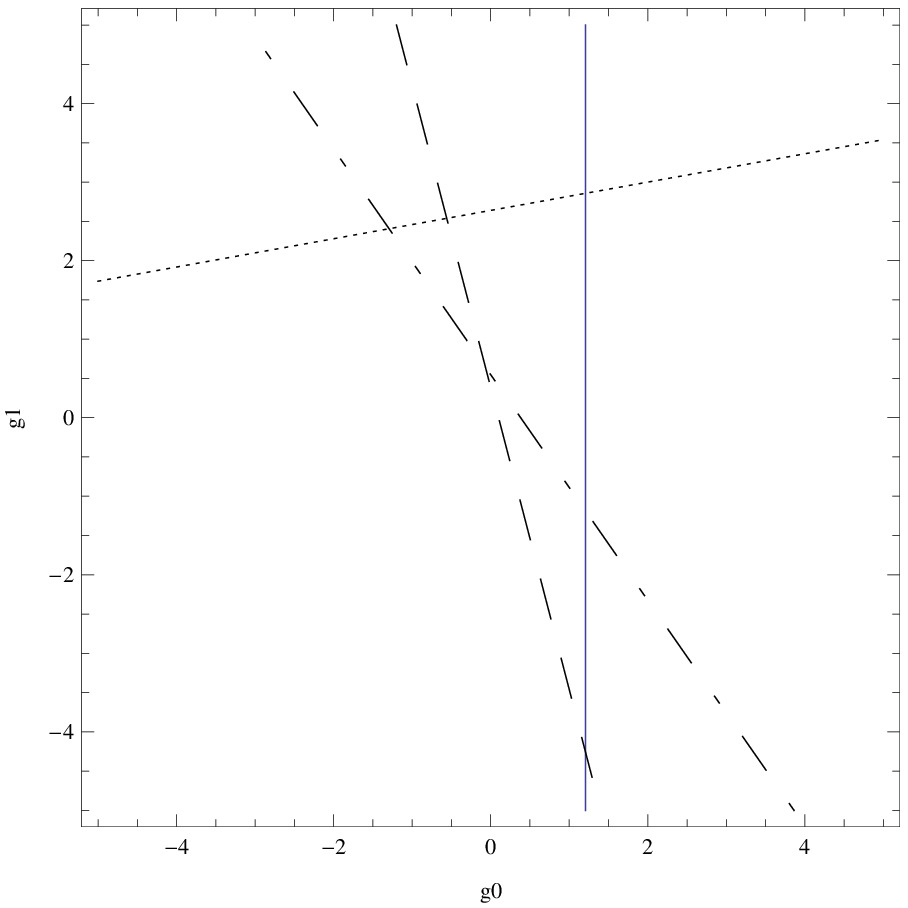}
\caption{The dependence of neutron EDM (solid), $^3He$ EDM (doted-dashed), $^3H$ EDM (doted) and parameter $\lambda$ on TRIV $\pi$-mesons iso-scalar and iso-vector coupling constants.}
\label{iso01}
\end{figure}

Another important feature of these experiments is the complementarity to other searchers for TRIV. To illustrate this we use results of the calculations of neutron and proton EDMs \cite{Liu:2004tq} and EDMs of few body nuclei \cite{Gudkov:2013edm} presented in terms of TRIV  meson-nucleon coupling constants. Then, assuming that TRIV pion, rho, eta, and omega meson coupling constants have about the same order of magnitude, we can write the main contributions to these EDMs in $e\cdot fm$ units as
\begin{eqnarray}
  d_n &\simeq & 0.14(\bar{g}^{(0)}_{\pi} -\bar{g}^{(2)}_{\pi}),\\
  d_p &\simeq & 0.14 \bar{g}^{(2)}_{\pi}, \\
  d_D &\simeq & 0.22 \bar{g}^{(1)}_{\pi}, \\
  d_{^3He} &\simeq & 0.2 \bar{g}^{(0)}_{\pi}+0.14 \bar{g}^{(1)}_{\pi}, \\
   d_{^3H} &\simeq & 0.22 \bar{g}^{(0)}_{\pi}-0.14 \bar{g}^{(1)}_{\pi},
\end{eqnarray}
where $\bar{g}^{(T)}_{\pi}$ is pion-nucleon TRIV coupling constant with isospin $T$. The comparison these results with eq.(\ref{eq:PTP}) shows that all these observable have different sensitivities to the models of TRIV. This became even more pronounced if we relax the assumption about values of TRIV coupling constants. These  sensitivities of  TRIV neutron scattering effect and neutron and  light nuclei  to TRIV $\pi$-mesons coupling constants are shown figures \ref{iso02} and \ref{iso01}. Therefore, one can see that even for the simplest case with the dominance of TRIV pion-nucleon coupling constants, it is necessary to measure at least three independent TRIV effects to constrain the source of CP-violation.

\section{Conclusions}

 We presented the summary of theoretical description of the TRIV effects in neutron transmission through a nuclei target and demonstrated that  these TRIV observables are free from FSI, and, as a consequence, are of the same quality as the EDM experiments. These  effects are enhanced by about $10^6$ due to the nuclear enhancement factor. In addition to this enhancement, the sensitivity to TRIV interactions in these effects  might be  enhanced by about $10^2$ if PV $\pi$-nucleon coupling constant is less than  the ``best value'' DDH estimate and by choosing right target.

	The main result of this paper is the proof of TRIV transmission theorem showing that the transmission of neutrons through an apparatus with arbitrary spin-dependent interactions that arise from time-reversal-invariant interactions is unchanged when the signs of all classical fields that interact with the neutron spin are reversed. We have used this result to construct tests of time-reversal invariance that are free of false asymmetries arising from combinations of time-reversal-invariant interactions and asymmetries in the apparatus.
 These types of experiments with high intensity neutron sources have a discovery potential of about $10^2 - 10^4$ for the improvement of the current limits on the TRIV interaction obtained from the EDM experiments.

\begin{acknowledgments}
This work was supported by the Department of Energy Grant No. DE-FG02-09ER41621 and by the Joint Institute of Nuclear Physics and Applications at Oak Ridge, TN.
\end{acknowledgments}

\bibliography{TViolation,ParityViolation}

\end{document}